\documentclass{ws-ijmpd}
\usepackage{url}
\usepackage[super,compress]{cite}
\begin{document}

\newcommand{\mnras}{MNRAS }
\newcommand{\apjl}{ApJ }
\newcommand{\apj}{ApJ }
\newcommand{\apjs}{ApJ }
\newcommand{\aj}{AJ }
\newcommand{\jcap}{JCAP}
\newcommand{\nat}{Nature }
\newcommand{\aap}{A\&A }
\newcommand{\aaps}{A\&AS }
\newcommand{\na}{New Astron. }
\newcommand{\pre}{Phys. Rev. E }
\newcommand{\prd}{Phys. Rev. D }
\newcommand{\prc}{Phys. Rev. C }
\newcommand{\physrep}{Physics Reports }
\newcommand{\aplett}{Astrophysical Letters }
\newcommand{\apss}{Astrophysics and Space Science }
\newcommand{\araa}{Annual Review of Astronomy and Astrophysics}

\newcommand{\gae}{\lower 2pt \hbox{$\, \buildrel {\scriptstyle >}\over {\scriptstyle
\sim}\,$}}
\newcommand{\lae}{\lower 2pt \hbox{$\, \buildrel {\scriptstyle <}\over {\scriptstyle
\sim}\,$}}
\newcommand{\aprop}{\lower 2pt \hbox{$\, \buildrel {\scriptstyle \propto}\over 
   {\scriptstyle \sim}\,$}}

\markboth{Zhang}
{GRB Prompt Emission}

%
\catchline{}{}{}{}{}
%

\title{Gamma-Ray Burst Prompt Emission}

\author{BING ZHANG}

\address{Department of Physics and Astronomy, University of Nevada,
Las Vegas, NV 89012, USA \\
zhang@physics.unlv.edu}



\maketitle

\begin{history}
\received{Day Month Year}
\revised{Day Month Year}
\end{history}

\begin{abstract}
The origin of gamma-ray burst (GRB) prompt emission, bursts of 
$\gamma$-rays lasting from shorter than one second to thousands of 
seconds, remains not fully understood after more than 40 years of 
observations. The uncertainties lie in several open questions
in the GRB physics, including jet composition, energy dissipation 
mechanism, particle acceleration mechanism, and radiation mechanism.
Recent broad-band observations of prompt emission with Fermi 
sharpen the debates in these areas, which stimulated intense
theoretical investigations invoking very different ideas. I will
review these debates, and argue that the current data suggest
the following picture: A quasi-thermal spectral component originating
from the photosphere of the relativistic ejecta has been detected
in some GRBs. Even though in some cases (e.g. GRB 090902B) this
component dominates the spectrum, in most GRBs, this component
either forms a sub-dominant ``shoulder'' spectral component in the low
energy spectral regime of the more dominant ``Band'' component, or 
is not detectable at all. The main ``Band'' spectral component 
likely originates from the optically thin region due to synchrotron
radiation. The diverse magnetization in the GRB central engine is
likely the origin of the observed diverse prompt emission properties
among bursts.
\end{abstract}


\ccode{PACS numbers:}


\section{GRB prompt emission}

Although discovered much earlier than the afterglow, the emission of 
gamma-ray bursts (GRBs) themselves, usually called the GRB 
{\em prompt emission}, is still poorly understood. The main uncertainties 
lie in several open questions in the GRB physics \cite{zhang11cr}: 
\begin{itemize}
 \item What is the composition of the GRB jet?
 \item What is the energy dissipation mechanism in the jet?
 \item Where is the region of energy dissipation and prompt emission?
 \item What is the particle acceleration mechanism in the energy dissipation region?
 \item What is the radiation mechanism of the particles?
\end{itemize}
Among these, the answer to the first question is more fundamental, which 
decides or at least strongly affects the answers to the rest of the 
questions. 

The traditional ``fireball'' model envisages a thermally-driven
explosion, with gravitational energy released during a catastrophic 
event (such as massive star core collapse or merger of two compact
stars) being deposited in the form of thermal heat at the base of 
the central engine. This fireball then expands under its own
thermal pressure and gets accelerated to a relativistic speed, if
baryon contamination in the ejecta is low enough
\cite{paczynski86,goodman86}. Most of the thermal energy is 
converted to the kinetic energy of the outflow \cite{shemi90},
with some thermal energy released as photons at the photosphere
\cite{meszarosrees00}. The kinetic energy of the outflow is 
further dissipated in the internal shocks \cite{rees94} and in the
external (forward and reverse) 
shocks as the ejecta is decelerated by the ambient medium
\cite{rees92,meszarosrees93}. The former powers the GRB prompt
emission, while the latter powers the afterglow 
\cite{meszarosrees97,sari98}. 

In contrast to the matter-dominated fireball picture, an opposite
opinion is that the entire GRB phenomenology is electromagnetic.
This is the electromagnetic (EM) model of GRBs
\cite{lyutikov03}. Here the magnetic field plays the dominant role in
ejecta dynamics. Little baryons are loaded from the central engine, so
that an extremely high value of $\sigma$ (ratio between Poynting flux 
and matter flux, or ratio between comoving magnetic energy density
and rest mass energy density) is invoked. It is conjectured that
$\sigma$ remains extremely high ($\sim 10^5 - 10^6$) at the deceleration
radius, so that the ``sub-Alfvenic'' condition $\sigma > \Gamma^2-1$ 
($\Gamma$ is the bulk Lorentz factor of the outflow) is satisfied.
Within this scenario, the magnetic energy is directly converted to
the particle energy and radiation energy through current instability
around the deceleration radius. Such a scenario requires that the
ejecta maintains high magnetization throughout, which is very 
difficult to achieve in reality.

More realistically, the GRB ejecta may carry both a matter component and a
magnetic component. This results in a wide range of magnetohydrodynamic (MHD) 
models of GRBs \cite{spruit01,drenkhahn02,giannios08,zhangyan11,mckinney12}. 
A real question is how high the $\sigma$ parameter is. Since $\sigma$ is
a function of radius (magnetic acceleration and dissipation can 
constantly change $\sigma$ in the flow), a more relevant question
would be how high the $\sigma$ parameter is at the GRB emission site. 
The complication is that one is not sure about the location of the GRB 
prompt emission radius $R_{\rm GRB}$ (see next section).
The values of $\sigma(R_{\rm GRB})$ and $R_{\rm GRB}$ are mutually
dependent (and also depend on a third parameter, i.e.
$\sigma_0$ at the central engine).
Different GRB prompt emission models invoke different
combinations of these parameters. 

Observationally, the currently available information of GRB prompt
emission includes the following aspects. A successful model should
be able to interpret them.
\begin{itemize}
 \item GRB lightcurves are extremely diverse \cite{fishman95}. Some are
spiky, while some others are smooth. Some have multiple (sometimes well separated)
spiky episodes, while some others have one or two smooth pulses. Some are very
long, while some others are very short. At least some 
lightcurves can be decomposed as the superposition of a fast and a slow
component\cite{gao12}. 
 \item GRB spectra can be well characterized. For most GRBs, both the
time-integrated and time-resolved spectra can be described by a so-called
``Band'' function, i.e. a smoothly-joint broken power law function
characterized by three parameters: the low-energy and high-energy photon
indices $\alpha$ and $\beta$, as well as the peak energy $E_p$ in the
energy spectrum \cite{band93}. Recent Fermi LAT/GBM observations suggest
that such a component dominates in nearly 7 orders of magnitude in energy
in most GRBs \cite{abdo09a,zhang11}. On the other hand, growing evidence
suggests that this phenomenological function cannot fully account for the
observations, and two additional spectral components, i.e. a quasi-thermal 
bump and a high-energy power-law component, are needed to fit the 
time-resolved data of bright GRBs 
\cite{zhang11,ryde10,guiriec11,axelsson12,ackermann11}.
 \item The peak energy $E_p$ varies in a wide range, from as high as
15 MeV in GRB 110721A \cite{axelsson12} to below 10 keV in GRB 060218
\cite{campana06}. It can evolve rapidly within a same burst, and two
patterns can be identified: hard-to-soft evolution throughout a pulse
or $E_p$ intensity tracking \cite{lu10,lu12}.
 \item GRB polarization measurements have been made for some GRBs. Even
though not with very high significance, evidence of high linear polarization
degree for at least some bright GRBs has been collected \cite{yonetoku11,yonetoku12}.
\end{itemize}
The available data do not have ``smoking gun'' criteria to differentiate among
the different models. Nonetheless, some available data already brought useful
clues, serving as ``finger prints'' to diagnose the underlying physics of
GRB prompt emission.

\section{GRB prompt emission models}\label{GRBmodels}

Before discussing the open questions and debates, it is informative to list
various prompt emission models and their properties. In the following, we
first categorize the models based on the location of GRB emission,
i.e. the radius of emission from the central engine $R_{\rm GRB}$. Swift
observations of early X-ray afterglow suggest that the prompt emission site 
is ``internal'' (below the external shock where the jet is decelerated) 
\cite{zhang06}. The
emission site is therefore bracketed between the photosphere radius $R_{\rm ph}$
(below which photons are opaque) and the deceleration radius $R_{\rm dec}$
(above which the jet decelerates due to jet-medium interaction). In the
literature, several possible emission sites have been proposed. From small to
large, they are, respectively,
\begin{itemize}
 \item Photosphere radius $R_{\rm ph}$: This radius is defined by the optically
thin condition for electron Thomson scattering, i.e. the last scattering surface
of photons originally trapped in the ejecta. The radius depends on luminosity $L$,
Lorentz factor $\Gamma$, and the latitude with respect to the observer. 
For example, for a matter dominated outflow with photosphere radius above
the coasting radius and for an on-axis observer, one has \cite{meszarosrees00}
$R_{\rm ph} \sim L \sigma_{\rm T}/4\pi m_p c^3 \Gamma^3 \sim 
3.7 \times 10^{11}~{\rm cm}~ L_{52} \Gamma_{2.5}^{-3}$, where 
the convention $Q_m=Q/10^m$ has been used. Considering angle-dependence of the
optical depth, one gets a concave photosphere shape with respect to the
line of sight \cite{peer08,beloborodov11}.
 \item Internal shock radius $R_{\rm IS}$: For a characteristic central engine 
variability time $\delta t$ and Lorentz factor $\Gamma$, the internal shock
radius can be estimated as \cite{rees94} $R_{\rm IS} \sim \Gamma^2 c \delta t 
\sim 3\times 10^{13}~{\rm cm}~ \Gamma_{2.5}^2 \delta t_{-2}$. Since there is
a range of $\delta t$ in a GRB lightcurve, internal shocks can actually
spread in a wide range of radii.
 \item Fast reconnection switch radius $R_{\rm rec}$: If the outflow is 
magnetically dominated at small radii and the field configuration is 
{\em striped-wind-like}, McKinney \& Uzdensky \cite{mckinney12} argued that a
switch from the slow, collisional reconnection regime to the fast, collisionless
reconnection regime occurs at around a radius $R_{\rm rec} \sim 10^{13}
-10^{14}$ cm. The GRB emission is assumed to occur at this radius, which
is above the photosphere. 
 \item Internal-collision-induced magnetic reconnection and turbulence radius
$R_{\rm ICMART}$: If the outflow is magnetically dominated at small radii
and the field configuration is {\em helical}, Zhang \& Yan \cite{zhangyan11} argued
that rapid discharge of magnetic energy cannot be realized until the field
lines are distorted enough through repetitive collisions. This radius is 
at least the internal shock radius (where collisions start), and should have
a typical value $R_{\rm ICMART} \sim 10^{15}-10^{16}$ cm. Within this model,
the slow variability component is related to the central engine activity,
while the fast variability component is related to mini-jets due to 
relativistic turbulent reconnection \cite{zhangyan11}, see also 
\cite{lyutikov03,narayan09,kumarnarayan09}. So the emission radius may
be estimated as $R_{\rm ICMART} \sim \Gamma^2 c \delta T \sim 3\times 10^{15}
~{\rm cm}~ \Gamma_{2.5}^2 \delta T$, where $\delta T \sim 1$ s is the 
variability time scale for the ``slow'' variability component.
 \item Deceleration radius $R_{\rm dec}$: The deceleration radius $R_{dec}$
is defined by the condition that the momentum collected from the ambient 
medium is comparable to the momentum in the jet. For a constant density medium,
this is typically $R_{\rm dec} \sim (3 E/2\pi\Gamma^2 n m_p c^2)^{1/3}
\sim 6.8 \times 10^{16}~{\rm cm}~ E_{52}^{1/3} \Gamma_2^{-2/3} n^{-1/3}$.
The electromagnetic model of GRBs \cite{lyutikov03} invokes current 
instabilities at this radius to power the observed GRBs. This is the largest
radius allowed by the ``internal'' requirement of GRB prompt emission.
\end{itemize}

Keeping in mind of various emission radii, in the literature the following
GRB prompt emission models have been discussed. The requirements for 
$R_{\rm GRB}$ and $\sigma(R_{\rm GRB})$ for each model are summarized in
Table 1.
\begin{itemize}
 \item The fireball internal shock model 
\cite{rees94,kobayashi97,daigne98,meszarosrees00}: This is the standard
theoretical framework. There are two emission sites in the model: the 
photosphere and the internal shocks. The magnetization parameter is
$\sigma \ll 1$ in both locations. 
 \item The magnetized engine - internal shock model: 
Some authors interpret the observed GRB spectrum as synchrotron emission
from the internal shocks only \cite{bosnjak09,daigne11}. An underlying assumption
of such a model is that the photosphere emission is suppressed. This may
appeal to a magnetized central engine \cite{daigne02,zhangpeer09}. So the
requirement of this model is $\sigma (R_{\rm ph}) \gg 1$, and $\sigma(R_{\rm IS})
\ll 1$. Some authors claim that it is possible that at the central engine
 $\sigma_0 (R_{0}) \gg 1$, while at both the photosphere radius and internal
shocks one has $\sigma (R_{\rm ph}) \ll 1$, and $\sigma(R_{\rm IS}) \ll 1$
\cite{hascoet13}. Such a rapid magnetic acceleration has not been seen in
numerical simulations.
 \item The dissipative photosphere model: Many authors \cite{thompson94,rees05,thompson06,ghisellini07c,giannios08,beloborodov10,lazzati10,ioka10,lazzati11,meszarosrees11,veres12} 
argued that the observed GRB Band spectrum is simply the emission from the 
GRB photosphere. In order to reconcile the theoretically expected spectrum 
(quasi-thermal) 
and the data, one needs to introduce energy dissipation around the photosphere,
probably due to internal shocks, proton-neutron collisions, or magnetic
reconnections. Non-thermal electrons accelerated from the dissipative
photosphere upscatter the seed thermal photons to make it non-thermal.
The internal shock emission is neglected in this model. Several sub-categories
of dissipative photosphere models have been discussed in the literature:
  \begin{itemize}
   \item Collisional heating model\cite{beloborodov10}: Heating of
photosphere is maintained via nuclear and/or Coulomb collisions. In this model,
one has $\sigma (R_{\rm ph}) \ll 1$ (to facilitate Coulomb collisions).
   \item Magnetic dissipation model\cite{giannios08,meszarosrees11,veres12}: 
Heating is maintained
by magnetic reconnection near the photosphere. A striped-wind magnetic field 
configuration is required in order to facilitate rapid reconnection at such
a small radius. To make efficient magnetic heating, one needs to have
$\sigma(R_{\rm ph}) \sim 1$. If $\sigma(R_{\rm ph})$ is too high, the 
photosphere luminosity is suppressed by a factor $(1+\sigma(R_{\rm ph}))$
\cite{zhangmeszaros02c,daigne02,zhangpeer09}. On the other hand, if 
$\sigma(R_{\rm ph})$ is too low, the energy released via magnetic dissipation
is small compared with the jet luminosity, so that the heating effect is not
significant.
   \item Jet-envelope interaction model\cite{ghisellini07c,lazzati10,lazzati11}: 
For long-duration GRBs that
are believed to be associated with deaths of massive stars, a relativistic
jet needs to penetrate through the stellar envelope first. The interaction
between the jet and the stellar envelope inevitably introduces dissipation
(through processes such as Kelvin-Helmholtz instabilities or collimation
shocks). This dissipated
energy can heat up the photosphere. In order to make such a mechanism
effective, the photosphere cannot be too far away from the outer boundary 
of the stellar envelope. Otherwise photons would be quickly ``thermalized''
before reaching the photosphere.
  \end{itemize}
 \item The internal collision-induced magnetic reconnection and turbulence
(ICMART) model: This model \cite{zhangyan11} invokes a large
emission radius $R_{\rm ICMART} \gg R_{\rm ph}$. Once fast reconnection is
triggered, the magnetic dissipation proceeds in a run-away manner. The 
magnetization parameter $\sigma(R_{\rm ICMART}) \geq 1$ when ICMART starts,
and $\sigma(R_{\rm ICMART}) \leq 1$ when ICMART is completed.
 \item The magnetic reconnection switch model: This model \cite{mckinney12}
has $\sigma(R_{\rm rec}) \geq 1$ to begin with, with $\sigma$ continuously
reducing with radius.
 \item The current instability model: This model \cite{lyutikov03} invokes
$\sigma (R_{\rm dec}) \gg 1$. The $\sigma$ value remains large beyond the
dissipation radius.
\end{itemize}

\begin{table}[ph]
\tbl{$R_{\rm GRB}$ and $\sigma(R_{\rm GRB})$ for different models.}
{\begin{tabular}{@{}cccc@{}} \toprule
Model & $R_{\rm GRB}$ & $\sigma(R_{\rm GRB})$ & References \\ \colrule
Fireball internal shock model & $R_{\rm ph}$ and $R_{\rm IS}$ 
 & $\sigma(R_{\rm ph}) \ll 1$, $\sigma(R_{\rm IS}) \ll 1$ & \cite{rees94,meszarosrees00}\\
Magnetized engine internal shock model & $R_{\rm IS}$ & $\sigma(R_{\rm ph}) \gg 1$, 
$\sigma(R_{\rm IS}) \ll 1$ & \cite{bosnjak09,daigne11} \\
Collisional heating photosphere model & $R_{\rm ph}$ & $\sigma(R_{\rm ph}) \ll 1$
& \cite{beloborodov10} \\
Magnetic heating photosphere model & $R_{\rm ph}$ & $\sigma(R_{\rm ph}) \lae 1$
& \cite{giannios08} \\
ICMART model & $R_{\rm ICMART}$ &  $\sigma_{\rm ICMART} \geq 1 \rightarrow \leq 1$
& \cite{zhangyan11} \\
& & ($\sigma(R_{\rm ph}) \gg \sigma(R_{\rm IS}) \gg 1$) & \\
Reconnection switch model & $R_{\rm rec}$ & $\sigma(R_{\rm rec}) \geq 1$ & \cite{mckinney12} \\
Current instability model & $R_{\rm dec}$ & $\sigma(R_{\rm dec}) \gg 1$ & \cite{lyutikov03} \\
\botrule
\end{tabular} \label{ta1}}
\end{table}

\section{Evolution of $\sigma$ and $\Gamma$ as a function of $R$.}

To ease further discussion below, we first discuss how $\sigma$
(and $\Gamma$) may evolve with radius $R$. 

In general, a GRB jet launched from the central engine may have two components,
one ``hot'' component due to neutrino heating from the accretion disk or the proto
neutron star, and a ``cold'' component related to a magnetic Poynting flux
launched from the black hole or the neutron star\cite{lei13,metzger11}.
The central engine can be characterized by a parameter
\begin{equation}
 \mu_0 = \frac{L_w}{\dot M c^2} = \frac{L_h  + L_{c}}{\dot M c^2}
= \eta (1 + \sigma_0),
\end{equation}
where $L_h = \eta \dot M$, $L_c = L_{\rm P}$, and $L_w$ are the luminosities of 
the hot component, cold component ($L_{\rm P}$ is the Poynting flux luminosity), 
and the entire wind, respectively.
The parameter $\sigma_0$ is defined as
\begin{equation}
 \sigma_0 \equiv \frac{L_c}{L_h} = \frac{L_{\rm P}}{\eta \dot M c^2}.
\end{equation}
For a variable central engine, all the parameters are a function of $t$. For 
simplicity, we do not introduce this $t$-dependence, but focus on the 
$R$-dependence of all the parameters.

After escaping from the central engine, the jet undergoes thermal acceleration
and magnetic acceleration and gains bulk Lorentz factor. At any radius $R$, the
flow can be categorized by a parameter
\begin{equation}
 \mu (R) = \Gamma (R) \Theta (R) (1 + \sigma (R)),
\end{equation}
where $\Gamma$ is the bulk Lorentz factor, $\Theta$ is the total co-moving
energy per baryon ($\Theta-1$ is the thermal energy), 
and $\sigma$ is the ratio between comoving cold (magnetic)
and hot (matter) energy densities.
Considering a completely cold magnetized outflow, i.e. $\eta=1$, $\Theta = 1$, 
and assuming no magnetic dissipation along the way, one would have 
\begin{equation}
 \mu_0 = 1+\sigma_0 = \mu= \Gamma (1+\sigma) = {\rm const}.
\end{equation}
In reality, thermal energy is inevitable, especially when the magnetic energy
is dissipated. About half of the dissipated energy is trapped in the system
as thermal energy (the other half used to accelerate the flow). As a result, 
$\mu$ would decrease from the original $\mu_0$ when photons escape from the
system (e.g. at the photosphere and other dissipation radii above the
photosphere). In some models (e.g. ICMART \cite{zhangyan11}), 
magnetic dissipation proceeds in a 
run-away manner, so that $\sigma$ and $\mu$ can drop quickly around a certain 
radius $R$, and copious photons are released from the system to power the prompt
emission.

A thermally driven fireball has a simple $\Gamma$-evolution history 
\cite{meszaros93,piran93,kobayashi99}: Initially the Lorentz factor 
$\Gamma$ increases linearly with $R$ until reaching the maximum Lorentz factor 
essentially defined by central engine baryon loading. The Lorentz factor
then ``coasts'' to the maximum value, reduces at the internal shocks
(due to loss of radiation energy), and finally decreases smoothly as a 
power law beyond the deceleration radius.

A magnetically-driven jet undergoes a more complicated evolution history
\cite{komissarov09,lyubarsky10,tchekhovskoy10,granot11}: For a magnetically
dominated jet with initial magnetization $\sigma_0$ ($\eta=1$), the jet
would first undergo a rapid acceleration until reaching $R=R_0$ where
$\Gamma(R_0) = \sigma_0^{1/3}$ and $\sigma(R_0) = \sigma_0^{2/3}$, and 
then go through a very slow acceleration process. The fastest acceleration 
proceeds as $\Gamma \propto R^{1/3}$, either via continuous magnetic
dissipation \cite{drenkhahn02} or via an ``impulsive acceleration'' of 
a short pulse \cite{granot11}. Ideally, one reaches the maximum Lorentz
factor $\Gamma = \sigma_0$ at the coasting radius $R_{c} = R_0 
\sigma_0^{2}$. However, the jet may start to decelerate before reaching
the coasting radius if $\sigma_0$ is large enough so that $R_{c} > R_{\rm dec}$.
Also magnetic dissipation can reduce the final coasting $\Gamma$, since energy
is released as prompt emission.

The $\Gamma$-evolution of a hybrid system (with both thermal energy and
magnetic energy at the central engine) is more complicated, and has not been
studied carefully in the literature. Since thermal acceleration proceeds 
more rapidly, it would be reasonable to assume that the thermal energy
gets converted to kinetic energy first, after which additional acceleration
proceeds magnetically if $\sigma_0$ is large enough.

\section{Debate I: is the jet strongly magnetized at $R_{\rm GRB}$?}

This is effectively to ask: how high is $\sigma(R_{\rm GRB})$? The two opponents 
in the debate and their corresponding arguments are summarized as follows:
\begin{itemize}
 \item Low $\sigma(R_{\rm GRB})$ (fireball internal shock model; magnetized engine
internal shock model, collisional heating photosphere model 
\cite{rees94,meszarosrees00,bosnjak09,daigne11,beloborodov10,ioka10}:
  \begin{itemize}
   \item The low-$\sigma$ model is the simplest, and seems to work reasonably well;
   \item We know the physics well, and give more robust predictions;
   \item Internal shocks and nuclear collisions are naturally expected in a GRB fireball;
   \item Even if the central engine is magnetized, one could have fast magnetic
acceleration so that $\sigma$ is brought below unity so that internal shocks can 
dissipate the kinetic energy.
  \end{itemize}
 \item Moderate to high $\sigma(R_{\rm GRB})$ (magnetic heating photosphere model,
ICMART model, reconnection switch model, current instability model
\cite{drenkhahn02,giannios08,zhangyan11,mckinney12,lyutikov03}: 
  \begin{itemize}
   \item The broad-band spectra of GRB 080916C show a dominant Band-component covering
6-7 orders of magnitude. Assuming this is the internal shock component, the predicted
photosphere component outshines the observed emission \cite{zhangpeer09,fan10}. This
suggests that the simplest fireball internal shock model does not explain the data
at least for this GRB. This would require a magnetized jet model with suppressed 
photosphere emission \cite{zhangpeer09}. A good candidate is the ICMART model
\cite{zhangyan11}.
   \item One way to overcome this difficulty is to introduce a magnetized central 
engine to suppress the photosphere emission, but demands that the magnetic energy
is converted to kinetic energy before internal shocks take place. However, 
various studies suggest that magnetic acceleration is not an efficient process
\cite{komissarov09,lyubarsky10,tchekhovskoy10,granot11}. It is essentially impossible 
to have $\sigma(R_{\rm ph}) \gg 1$, while $\sigma(R_{\rm IS}) \ll 1$. 
   \item The internal shock model itself suffers a list of difficulties 
(see e.g. \cite{zhangyan11} for a summary): low radiative efficiency, fast cooling,
too many electrons to interpret $E_p$, not easy to interpret the $E_p \propto
E_{\rm iso}^{1/2}$ \cite{amati02} and $E_p \propto L_{\rm iso}^{1/2}$ 
\cite{yonetoku04} relations. The difficulties can be overcome if the flow
is magnetically dominated and dissipation is through the ICMART process.
  \end{itemize}
\end{itemize}

\section{Debate II: Is the jet energy dissipated via shocks or magnetic reconnection?}

This question is closely related to Debate I. If indeed one can have $\sigma(R_{\rm IS})
\ll 1$, then internal shocks would accelerate electrons to the desired energy to
power GRB prompt emission. Recent particle-in-cell (PIC) simulations have shown
that magnetic relativistic shocks are not good at accelerating electrons
\cite{sironi09a,sironi11a}. For an electron-ion plasma, efficient electron 
acceleration is possible only when $\sigma \lae 10^{-3}$ for relativistic shocks
\cite{sironi11a}. For mildly relativistic shocks the constraint is less stringent
\cite{sironi09a}. In any case, in order to make internal shocks efficient 
accelerators, one has double constraints: a not-too-high $\sigma$ and 
a not-too-strong shock. The latter constraint further limits the already 
low dissipation efficiency of shocks.

If $\sigma \gae 1$, the particle acceleration mechanism is likely through
magnetic reconnection. In this case, internal shocks can help to facilitate
rapid reconnection \cite{zhangyan11,sironi11b}.

\section{Debate III: Is the dominant radiation mechanism quasi-thermal or non-thermal?}

This question is related to the origin of the ``Band'' spectral component
observed in most GRBs, which is dominant spectral component of GRB
prompt emission.
The two leading interpretations invoke completely different radiation mechanisms.
One scenario interprets the spectrum as upscattered quasi-thermal emission from
the photosphere, with $E_p$ essentially defined by the photosphere temperature.
The other scenario interprets the spectrum as non-thermal emission in the
optically thin region. The arguments for each of the above two scenarios include
the following:
\begin{itemize}
 \item Quasi-thermal (dissipative photosphere models 
\cite{thompson94,rees05,peer08,beloborodov10,giannios08,lazzati10,ioka10}): 
This model interprets the Band component as modified thermal emission from
the photosphere. The broadening mechanism includes both physical broadening
(Compton upscattering of the seed thermal photons) and geometric broadening
(the effect of equal-arrival-time volume) \cite{peer08,beloborodov11,lundman13}.
The arguments that have been raised to support this scenario include:
  \begin{itemize}
   \item The observed $E_p$ distribution among GRBs is narrow, all around MeV range 
and below. The photosphere temperature is in this range, and does not sensitively
depend on luminosity \cite{beloborodov13};
   \item The ``broadness'' of the GRB spectra around $E_p$, which can be described
by a Band function, is relatively ``narrow'' (compared with the synchrotron peak
of other objects, such as blazars). So it is likely of a thermal origin 
\cite{beloborodov13}.
   \item The photosphere model can interpret \cite{fan12} the Amati relation 
\cite{thompson07} and other correlations, including a correlation between $\Gamma$
and $E_{\rm iso}$ or $L_{\rm iso}$ \cite{liang10,ghirlanda12,lv12}.
   \item The photosphere emission has a high efficiency. It is hard to avoid
a bright photosphere.
  \end{itemize}
 \item Non-thermal (e.g. internal shock model, ICMART model, 
\cite{meszaros94,tavani96,lloyd00,bosnjak09,daigne11,zhangyan11}): This model
interprets the Band component as non-thermal emission of electrons, such as
synchrotron or synchrotron self-Compton. The arguments in support of this 
scenario include the following:
  \begin{itemize}
   \item The observed $E_p$ distribution among GRBs is actually not that narrow.
Considering X-ray flashes (the softer GRBs), $E_p$ can distribute from above
MeV to the keV range. More importantly, $E_p$ can evolve rapidly within a same
GRB. One can define a ``death line'' in the $L-E_p$ domain for the dissipative 
photosphere model that interprets $E_p$ as the temperature of the outflow 
\cite{zhang12}. GRB 110721A shows a very high $E_p \sim 15$ MeV early on
\cite{axelsson12}, which is well above the ``death line'' \cite{zhang12}.
This rules out the thermal origin of $E_p$ at least for this burst
(see also \cite{veres12b}). Since
the Band function parameters are typical among other GRBs, this would suggest
that the Band component in most other GRBs may be also of a non-thermal 
(e.g. synchrotron) origin.
   \item Due to a small emission radius, the GRB spectrum predicted in the 
photosphere model cannot extend to above GeV (due to photon-photon pair
production opacity). In some GRBs (e.g. GRB 080916C), the Band component 
extends to much higher energies \cite{abdo09a,zhang11}. One counter-argument
would be that the GeV emission is of an external shock 
origin\cite{kumar09,kumar10,ghisellini10}. However, it is now clear that
the GeV afterglow only sets in after the end of prompt emission 
\cite{zhang11,maxham11,he11,liu11}, and during the prompt emission phase,
the GeV emission tracks the variability in the sub MeV regime and should
have an internal origin.
   \item A major difficulty of the photosphere model is that it predicts
too hard a low-energy spectrum (with $\alpha \sim +0.4$\cite{beloborodov10,deng13})
to match the observed $\alpha = -1$ spectrum. Several efforts to soften the
spectrum has been made. 1. Synchrotron radiation is introduced to contribute
to emission below $E_p$. However, the predicted spectrum \cite{vurm11} does
not have the correct shape to interpret the data. In particular, for a softer
low energy spectrum, the high energy spectrum is also softened, a behavior
not observed from the data. 2. If the high-latitude emission dominates
the spectrum, due to the geometric
broadening effect one can get $\alpha = -1$ \cite{peer11}. However,
during the prompt emission phase, new materials are continuously ejected,
whose emission outshines the high-latitude contribution. The predicted 
spectrum is still too hard to interpret the data 
during most of the prompt emission phase. 
3. By introducing a special form of angular structured of the jet, one
can reproduce $\alpha = -1$ \cite{lundman13}. The special form of the
structured jet requires that the luminosity keeps constant in
a wide range of angle, while the bulk Lorentz factor drops with angle
as a power law.  Such a jet structure is most helpful to enhance the high-latitude
emission, since a low $\Gamma$ from the wing allows a wide $\Gamma^{-1}$
cone to contribute to the on-axis observer. The high-luminosity in the
wing (high latitude) then significantly contributes to the observed emission and
therefore softens the spectrum. It is however not known how typical such 
a jet structure is. If both luminosity and Lorentz factor have an angular 
structure, the photosphere spectrum would then not so different from the
uniform jet case, which predicts a harder spectrum.
   \item The dissipative photosphere models require some tuning of parameters.
For example, the magnetic dissipation model requires
$\sigma (R_{\rm ph})$ to be around unity to assure a high radiation efficiency; 
and the requirement of thermalization poses stringent constraints on the
value of Lorentz factor \cite{vurm13}.
   \item Three independent clues (GeV extension, X-ray tail, and optical
association) all point towards a large emission radius $R_{\rm GRB}$ for 
GRBs \cite{kumar07,gupta08,zhangpeer09,zhangyan11,hascoet12}. Although for 
each argument one has to make the assumption that the sub-MeV emission comes 
from the same region as the emission from the other wavelength, when
combining all three criteria, it is 
rather unlikely that only sub-MeV emission comes from the photosphere
while emission from other three bands are all from a larger radius.
   \item Synchrotron models also predict $E_p \propto L^{1/2}$ given
that the emission radius $R_{\rm GRB}$ is not very different from burst
to burst \cite{zhangmeszaros02c}. This may be satisfied for the ICMART
model \cite{zhangyan11}. Other correlations \cite{liang10,ghirlanda11,lv12}
can be interpreted with a high-$\sigma$ central engine model \cite{lei13}
that invokes the Blandford-Znajek mechanism \cite{blandford77}. 
   \item Very rapid hard-to-soft $E_p$ evolution during the rising phase
of the leading GRB pulse \cite{lu10,lu12} could be a big challenge to the
photosphere model, but can be interpreted within the framework of
synchrotron radiation model \cite{zhangyan11,uhmzhang13}.
   \item In several GRBs a quasi-thermal component is found in superposition
with the Band component \cite{ryde05,guiriec11,axelsson12,guiriec13},
whose temporal evolution is consistent with the photosphere origin 
\cite{ryde09}. If this component is of the photosphere origin, then the
main Band component must be another component. Since the Band component
extends to below the quasi-thermal component, it must not be self-absorbed
in the emission region, which means $R_{\rm Band} \gg R_{\rm ph}$, i.e.
the Band component comes from an optically-thin region far above the
photosphere.
  \end{itemize}
\end{itemize}

\section{A breakthrough: fast cooling synchrotron radiation in a decaying
magnetic field as the origin of the Band spectral component}

Two leading non-thermal radiation mechanisms to interpret GRB prompt 
emission are synchrotron radiation \cite{meszaros94,tavani96,lloyd00} and
synchrotron self-Compton (SSC). The SSC mechanism was particularly suggested
to interpret the prompt optical emission of the ``naked-eye'' GRB 080319B,
whose flux significantly exceeds the low-energy extrapolation of the 
gamma-ray emission \cite{racusin08,kumarpanaitescu08}. Further studies
suggest that this mechanism is disfavored due the following arguments:
1. A dominant SSC component usually predicts an even more dominant 2nd-order
SSC component, which greatly adds to the total energy budget of GRBs
\cite{derishev01,piran09}; 2. This mechanism predicts a strong optical 
flash. However, observations show that the case of GRB 080319B is rare.
Most prompt optical flux is consistent with or below the low-energy
extension of gamma-rays \cite{shen09}; 3. The gamma-ray lightcurve of
GRB 080319B is much more variable than the optical lightcurve. 
Simulations suggest that the SSC and synchrotron lightcurves show 
show similar degree of variability \cite{resmi12}. Besides these
criticisms, in general, the SSC mechanism predicts a broader $E_p$
distribution than the synchrotron mechanism due to its sensitive 
dependence on the electron Lorentz factor $\gamma_e$ (4th power
as compared with 2nd power for the synchrotron mechanism) 
\cite{zhangmeszaros02c}.

The synchrotron mechanism is known to power non-thermal emission of
many other astrophysical phenomena (e.g. blazars, micro-quasars,
supernova remnants, pulsar-wind nebulae). Within the GRB context,
the main criticism has been the fast-cooling problem. Since the
magnetic field strength is strong in the GRB prompt emission region,
electrons lose energy in a time scale much shorter than the dynamical
time scale. Traditionally, it was believed that the spectral index
below the injection energy due to fast cooling gives a photon
index -1.5, too soft to account for the typical value $\alpha \sim -1$
\cite{sari98}. This has been regarded as a main criticism against
the synchrotron mechanism \cite{ghisellini00}. 

Recently, a breakthrough was made in modeling synchrotron radiation
in GRBs. By introducing a decrease of magnetic field with radius
(e.g. due to flux conservation in a conical jet, probably with
additional magnetic dissipation), a natural physical ingredient 
previously ignored in modeling synchrotron cooling, 
Uhm \& Zhang\cite{uhmzhang13} found that a
low-energy spectrum of the fast-cooling synchrotron spectrum
is typically harder than -1.5. By varying model parameters
(e.g. the decay index $b$ and the magnetic field normalization $B_0$
at radius $r_0 = 10^{15}$ cm from the central engine, convention
$B(r) =B_0 (r/r_0)^{-b}$), one can reproduce a range of $\alpha$
centered around $\alpha \sim -1$, with a distribution from
$\alpha \sim -1.5$ to $\alpha > -0.8$. 
More interestingly, within the typical bandpass of GRB detectors,
the synchrotron model spectrum is close to the phenomenological
Band function (Fig.1). As a result, we suggest that fast-cooling
synchrotron is likely the dominant mechanism that shapes the
``Band'' component of the observed GRB spectra.

\begin{figure}[htb]
        \centering\leavevmode\epsfxsize=3truein
        \psfig{figure=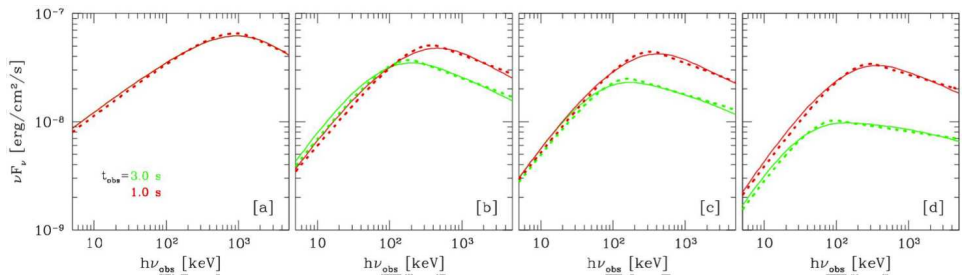,width=5in}  
        \centering
       \parbox{5in}{
        \caption{{\small The model spectra (solid lines) of 
fast-cooling synchrotron spectrum with a decreasing magnetic field in 
the emission region, as is naturally expected in the GRB environment. 
The Band function fits (dotted lines) are over-plotted for comparison.
One can see that the fast-cooling synchrotron spectra can mimic the Band
function and may provide a reasonable fit to the GRB spectral data.
From Uhm \& Zhang (2014)\cite{uhmzhang13}.}
       }
        \label{fig:syn-Band}
}
\end{figure}

\section{GeV emission}

Fermi LAT detected emission in the GeV range from a sample of bright GRBs. Two
important features of the GeV emission is that some of them have a delayed onset
and the emission typically decays as a power law \cite{abdo09a,abdo09b,zhang11}.
This led to the suggestion that the GeV emission is from the external shock
\cite{kumar09,kumar10,ghisellini10}. Further studies suggest that the external
shock interpretation applies to phase when the MeV prompt emission is over.
During the prompt emission phase, the GeV emission should still have an
internal origin \cite{maxham11,he11,liu11}.

Several suggestions have been proposed to interpret the delayed onset, e.g.
delayed onset of the hadronic component \cite{razzaque10,asano12} and delayed onset 
of IC scattering off the cocoon photons \cite{toma09}. On the other hand,
the data can be understood in terms of gradual decrease of pair-production
opacity \cite{zhang11}, which may be achieved in a scenario invoking slow
acceleration of a magnetized jet \cite{meszarosrees11,bosnjak12}.

A hadronic origin of prompt emission has been discussed in the literature
\cite{asano09b,asano12}. In order to have a dominant hadronic emission component,
the proton-to-electron energy ratio has to be large \cite{gupta07b}, which
greatly increases the energy budget of the bursts. Such a high proton fraction
is already constrained by the non-detection of neutrinos from GRBs by IceCube
\cite{icecube11,icecube12,he12,zhangkumar13}.

\section{Putting pieces together}

Combining insights gained from multi-wavelength data and theoretical modeling,
one may draw the following global picture regarding the origin of GRB prompt 
emission:
\begin{itemize}
 \item The Band spectral component originates from an optically thin region
with a non-thermal emission mechanism, very likely synchrotron radiation. 
Fast cooling synchrotron radiation in a decreasing magnetic field gives
a simple picture to interpret the Band component, including its typical
-1 low energy photon index \cite{uhmzhang13}. Alternatively, one may introduce 
slow cooling \cite{peerzhang06} or slow heating due to magnetic turbulence 
\cite{zhangyan11,asano09} to achieve the right low-energy spectral index.
 \item The photosphere component has been observed: it can dominate the spectrum
in rare cases such as GRB 090902B \cite{ryde10,zhang11}, but mostly show up as 
a superposed bump in the low-energy wing of the Band component \cite{guiriec11,axelsson12},
or is essentially suppressed (e.g. in GRB 080916C \cite{abdo09a,zhang11,zhangpeer09}).
 \item The diverse spectral behavior of GRB prompt emission may be related
to different $\sigma_0$ values at the central engine \cite{zhang12}. 
If $\sigma_0$ is small enough, one can have
a dominant photosphere spectrum as observed in GRB 090902B \cite{ryde10,zhang11}.
If $\sigma_0$ is moderately high, the photosphere emission is not completely suppressed,
and can have contribution to the observed spectra along with the non-thermal emission
from the optically thin region. This applies to the cases such as GRB 100724B \cite{guiriec11}
and GRB 110721A \cite{axelsson12}. Finally, if $\sigma_0$ is high enough, the photosphere
emission is greatly suppressed, and the observed spectrum is dominated by the non-thermal
Band component arising from the magnetic dissipation (e.g. ICMART) region.
\end{itemize}

\section{Other clues, future prospects}

So far the observational clues are limited to the spectra and lightcurves of GRB
prompt emission. Other observational clues can help to further test and constrain the models.

Photon polarization is essential to diagnose jet composition, geometric configuration,
and radiation mechanism of GRB jets \cite{toma09b}. Recently, $>3\sigma$ polarized
gamma-ray signals have been detected in 3 bright GRBs \cite{yonetoku11,yonetoku12}.
The degree of polarization is typically very high (several tens of percent). This immediately
disfavors the internal shock model that invokes random magnetic fields in the shocked
region (due to plasma instabilities) and the non-structured low-$\sigma$ photosphere
models. Models invoking dissipation of ordered magnetic fields, e.g. 
\cite{zhangyan11,mckinney12,lyutikov03}, are favored. A structured jet photosphere
model may also produce polarized photons via Compton scattering, but the polarization
degree would have a different energy-dependence from the synchrotron model in
ordered magnetic fields. A high-sensitivity gamma-ray polarimeter with a wide band-pass
to detect energy-dependent polarization signals is essential to constrain these models.

The IceCube Collaboration has placed a more and more stringent high-energy neutrino
flux from GRBs \cite{icecube11,icecube12}. The upper limit starts to challenge the conventional
internal shock model \cite{he12}. 
A model-dependent study of the GRB neutrino models \cite{zhangkumar13}
suggests that the neutrino signal would be stronger in the photosphere models than the 
internal shock model given a same cosmic-ray-to-gamma-ray flux ratio 
(see also \cite{gaoshan12}). If neutrinos are still not detected from GRBs
in the next several years, then the dissipative photosphere models that invoke 
proton acceleration would be further constrained. Models invoking magnetic dissipation
at a large radius \cite{zhangyan11} are well consistent with the neutrino flux 
upper limit constraint, and would be supported if neutrinos continue not to be 
detected to accompany GRBs in the future.


 I acknowledge recent stimulative collaborations with Z. L. Uhm, B.-B. Zhang, 
P. Kumar, A. Pe'er, H. Yan, W. Deng, L. Resmi on the subject of GRB prompt 
emission. I also thank interesting discussion with the following people
on the subject: K. Asano, A. Beloborodov, Z.-G. Dai, F. Daigne, Y.-Z. Fan, 
H. Gao, D. Giannios, K. Ioka, S. Kobayashi, D. Lazzati, Z. Li, E.-W. Liang, 
P. M\'esz\'aros, R. Mochkovitch, K. Murase, S, Razzaque, F. Ryde, X.-Y. Wang, 
and X.-F. Wu. This work is supported by NSF AST-0908362.




\end{document}